\documentstyle[aps,graphicx]{revtex}
\begin{document}
\title{Unusual bound or localized states}
\author{M. A. Cirone$^{a},\; $G. Metikas$^{a}$
and W. P. Schleich$^{a,b}$}
\address{$^{a}$ Abteilung f\"{u}r Quantenphysik,
Universit\"{a}t Ulm, D-89069 Ulm, Germany \\
$^{b}$ Department of Physics, North Texas State University,
Denton, TX \\}

\maketitle

\begin{abstract}
We summarize unusual bound or localized states in quantum mechanics.
Our guide through these intriguing phenomena is the classical physics
of the upside--down pendulum, taking advantage of the analogy
between the corresponding Newton's equation of motion and the
time independent Schr\"{o}dinger equation. We discuss the zero--energy
ground state in a three--dimensional, spatially oscillating, potential.
Moreover, we focus on the effect of the attractive quantum
anti--centrifugal potential that only occurs in a two--dimensional
situation.
\end{abstract}
\vspace{2.0cm}

\draft

\twocolumn

\section*{1. The pendulum: a guide to quantum physics}

Determining the energy levels of a quantum system was a desperate
enterprise before the advent of the Bohr--Sommerfeld quantization
conditions. Max Planck had emphasized that the energy of the harmonic
oscillator is quantized in units of a fundamental energy given by the
product of what we now call Dirac's constant and the frequency of
the oscillator. However, this rule failed miserably when used in other
quantum systems. Paul Ehrenfest's adiabatic principle applied to the
pendulum \cite{ehr} whose length slowly changes as a function of time,
made clear that it is not the energy, but the action that is quantized,
but why? When we change the length of the pendulum adiabatically, the
amplitude of oscillation does not stay constant, neither does the frequency
nor the energy. What stays constant is the action, that is, the area in
phase space. For the founding fathers of quantum mechanics it must have
been a miracle, an amazing fact symbolizing in post--Schr\"{o}dinger
language that the number of nodes in an energy wave function stays constant
under adiabatic changes.

The classical dynamics of the pendulum serves as an excellent guide for many
quantum phenomena. For example, the upside--down pendulum yields insight
into the energy wave function of a periodic potential \cite{whe}. Here we 
do not focus on the familiar Bloch states which appear when the potential 
enjoys a strict periodicity over the whole space. Our states occur when the
modulation of the potential extends only over a finite domain of space.
Due to the shape of this potential we refer to it as the {\em accordion
potential}. Extensions of these one--dimensional considerations to two and
three
 dimensions lead us to the effect of the quantum anti--centrifugal force
\cite{cir}.

In the present paper we take seriously the joke
`physics takes mathematics and makes it understandable', therefore we
shall suppress all the mathematics and highlight the essential ideas.
This approach is justified by the fact that we are not inventing or
applying new mathematics, but attempt to draw together phenomena
of different fields exposing a common thread. We shall focus on a general
point of view but emphasize that the modern tools of cold atoms in a
standing electromagnetic wave can demonstrate these unusual bound or
localized states predicted in this paper.

We follow one `Leitmotif', that is, a theme common to all phenomena
discussed in this article: The classical physics of the upside--down
pendulum. Moreover, we develop a little side theme summarized by the
phrase "wave mechanics in quantum potentials".

Our paper is organized as
follows:
In Sec.2 we briefly summarize the classical physics of the upside--down
pendulum and then turn in Sec.3 to related phenomena, such as the Paul trap,
the helium atom and field induced multipoles. We devote Sec.4 to the
transition to quantum mechanics and in particular, discuss the shape of
the ground state wave function in the accordion potential. In Sec.5 we
show that in three dimensions a localized ground state can exist
-- even at zero energy. The quantum anti--centrifugal potential emerges
when in Sec.6 we focus on the Schr\"{o}dinger equation in two
dimensions. We conclude in Sec.7 with an outlook and present extensions
of the concepts discussed in this paper.
An Appendix summarizes the
mathematical aspects of the upside--down pendulum and the
Schr\"{o}dinger equation of the particle in a periodic potential.
 
\section*{2. The upside--down pendulum}

In the mid-forties it was recognized \cite{bet1,bet2} that the
cyclotron would never be able to
accelerate particles to energies above a critical value. The problem lies
in the fact that in free space static electric or magnetic or both fields
together cannot focus a beam of charged particles in all planes through
the axis of the beam. Fortunately, this difficulty, which applies to linear
accelerators, can be circumvented to some degree in circular systems.
However, the degree of focusing permitted by the effect of the centrifugal
force is extremely weak. These considerations suggested
that energies higher than 10 MeV are not attainable
without unreasonable expenditures of money and materials. However, in 1949
Nick Christofilos from Athens, who at that time was an elevator operator,
dreamt up the principle of strong focusing. Although he got a patent,
his work was unpublished and the principle was re--discovered \cite{cou}
in 1952.

The central idea of the principle of strong focusing is to use a sequence
of strong focusing and defocusing fields to achieve a net focusing effect.
This idea is best illustrated using the principle of the upside--down
pendulum \cite{lei}. Consider the motion of a point mass on one end of a
massless rod. The opposite end of the rod is connected via a hinge to another
rod that is fixed to a support. This pendulum experiences a constant
gravitational force pulling downwards. When the point of suspension is such
that the pendulum hangs down and we concentrate on the limit of small
displacements from the stable hanging position we can approximate the
motion of the mass by that of a harmonic oscillator. However, when the
pendulum is upside--down and stands straight against the gravitational force,
the motion is unstable and the mass tends to fall down. In this case
we face an inverted harmonic oscillator of negative steepness. In order
to stabilize it, Pyotr L. Kapitza \cite{lan} in 1951 suggested a rapid
vertical modulation of the foundation of the pendulum. When the modulation
frequency is above a critical value determined by the length of the pendulum,
the gravitational acceleration and the modulation amplitude, the
motion of the pendulum becomes stable. Why is this so?

Many mathematical arguments offer themselves. They range from the method
of averaging \cite{bog} via the secular growth theory \cite{kev} to the
Floquet theorem \cite{whi}. However, none of these mathematical techniques
provide deeper insight into the physics of the stabilizing mechanism.
For the sake of argument and
to illustrate this lack of insight, we follow the Floquet reasoning.

The equation of motion for the phase angle $ \varphi = \varphi (t)$ of the
pendulum, measured from its upside--down unstable position, is the Mathieu
equation \cite{abr}, as discussed in the Appendix.
The Floquet theorem \cite{whi} determines the domains of stability
and instability of this equation. Hence, it is the stability chart
\cite{abr} of the Mathieu equation that governs the parameter regime
for which the upside--down pendulum is stable.

What is the mechanism of this stabilization? When the foundation
accelerates the pendulum upwards, the gravitational force which pulls the
pendulum down and makes its motion unstable, is reduced. A push up therefore
corresponds to a stabilizing force. On the other hand, a motion of the
foundation downwards increases the effective gravitational force and thereby
enhances the instability of the pendulum: A pull down corresponds to
a de--stabilizing force. On the first sight one might think that during
a complete cycle of modulation, the two effects average out, however,
during a
majority of time there is a pushing up effect. Why is this?

In the stable mode the pendulum performs two types of motion, the secular
and the micro--motion. The secular motion corresponds to a slow
oscillation of the mass between two extremes of phase angles. In
addition, the mass undergoes a rapid motion with a frequency identical
to the modulation frequency. It is this micro--motion, superimposed on the
secular motion, which makes the pendulum stable and which breaks the symmetry
between the upward and the downward push. Indeed, due to the geometry
of the pendulum, the angular acceleration is not only determined by the
acceleration due to the modulation, but by the product of this acceleration
and the instantaneous phase angle. The corresponding force is therefore
a tidal force: No displacement of the oscillator --- no force; large
displacement --- large force. The modulation of the foundation translates into
a modulation of the phase angle. Since the effective acceleration is the
product of the acceleration due to the modulation and the phase angle we find
the product of two oscillatory functions with the same phase. The time
average of such a quantity is a positive constant providing a harmonic
oscillator potential of positive steepness for the slow motion.

\section*{3. Paul trap, helium atom and negative ions}

Many other applications of the upside--down pendulum come to mind, for
example the Paul trap \cite{pau} that allows us to store and manipulate single
ions in a controlled way. It is worth mentioning that Paul traps play an
important role in the recent proposals for a quantum computer.

In the Paul trap the need for a time dependent force is dictated by the
Poisson equation of electrostatics making it impossible to create binding
forces in all three dimensions of space. Indeed, the Poisson equation enforces
the feature that even when two spatial directions enjoy a binding potential
the third one is anti--binding. To overcome the instability of the trap in
the third direction two possibilities offer themselves: apply a time
independent homogeneous magnetic field that confines the ion in the
third direction or apply an alternating voltage to the trap. The first approach
corresponds to the Penning trap, the second to the Paul trap. In the most
elementary case the Paul trap consists of a quadrupole field, giving rise
to harmonic oscillator potentials. The alternating voltage applied to the
electrodes creates a dynamical binding in three dimensions. In each dimension
the physics of this binding phenomenon is identical to the upside--down
pendulum.

The helium atom represents another interesting application of this concept
of dynamical binding. For more than one hundred years physicists have
tried to understand the motion of two electrons around the nucleus. A
one--dimensional model in which the electrons move along a line with the
nucleus at the origin brings out the essential physics. On first sight,
a situation in which the electrons are on different sides of the nucleus
seems to be preferable since in this way the electrons can avoid each other
in a most effective way. However, an extremely interesting situation
\cite{ric} occurs when both electrons are on the same side. Indeed, here
one electron is close to the nucleus and oscillates rapidly between two
equilibrium points. The second electron is far away from the nucleus and
moves slowly feeling the attractive force of the nucleus and the repulsion
from the inner electron. Indeed, the fast motion of the inner electron creates
a time averaged repulsive potential, which superposes with the attractive
Coulomb potential of the nucleus, giving rise to a local potential minimum
for the outer electron. In this case, the electron--electron repulsion,
together with the rapid motion of the inner electron, forms a time dependent
barrier to stop the unstable motion of the outer electron caused by the
attraction towards the nucleus. This phenomenon is the upside--down
pendulum in disguise \cite{sai}.

Our last illustration of effective potentials, arising from time averaging,
is the effect of an induced dipole or multipole.  It helps us to make the
transition to unusual bound states in quantum mechanics. Two examples offer
themselves, ({\em i}) the binding of an electron in a negative ion and ({\em
ii}) the atom in a
 standing electromagnetic wave.  The field of an
electron can polarize a
 neutral atom even when the electron is at a distance
large compared with the
 atomic dimensions. This interaction between the
electron and the atom leads to
 a force of attraction.  This attraction is the
reason why some atoms, such as hydrogen or halogen, are able to form negative
ions by an attachment of an electron.   

We now consider the motion of an atom in a standing electromagnetic wave.
Indeed, in quantum optics many such experiments have been and are being
performed. The light induces a dipole moment in the atom and this dipole again
interacts with the light field. Since the interaction energy is the product of
the induced dipole and the field, it enters quadratically. Consequently, the
position dependence of the interaction energy follows from the position
dependence of the square of the field that is, from the square of the mode
function thereby creating the effective potential for the motion of the atom.

\section*{4. Energy ground state in the accordion potential}

All examples in the previous sections illustrate how dynamical binding can
occur in classical physics. We now turn our attention from Newton's equation 
of motion

\begin{equation}
\frac{d^2 \varphi(t)}{dt^2}+\left[\xi_{0}+\xi(t) \right]\varphi(t)=0
\end{equation}
for a harmonic oscillator with time dependent
steepness $\xi(t)$ to the time independent Schr\"{o}dinger equation 

\begin{equation}
\frac{d^2u(x)}{dx^2} + \frac{2M}{\hbar^2} \ [E - V(x)] \ u(x)= 0 
\label{Schrodinger}
\end{equation}

\noindent for a point particle of mass $M$ in a position dependent potential
$V(x)$; both equations are identical in form.
The role of the angle coordinate of the classical upside-down pendulum is now
played by the energy wave function $u$. Moreover, the time variable $t$ 
is replaced by the
coordinate variable $x$. Indeed, Newton's equation of motion is an 
equation of
second order in time for the position and the Schr\"{o}dinger equation is an
equation of second order in position for the wave function. We emphasize that
this analogy between the two equations only holds true for the
classical harmonic oscillator. Whereas the Schr\"{o}dinger equation is always
linear, a feature independent of the form of the potential, the form of
Newton's equation strongly depends on the shape of the potential and in
general is nonlinear.

We focus on a periodic potential in space.
The particular shape of this potential is of no importance as long as it is
averages out to zero over space; it is as often positive as it is negative.
For our analysis it is crucial that the oscillations of the
potential start at a given point in space and end thereby connecting a flat
space through a domain of wiggles back to a flat space.
We measure energy relative to flat space. Fig.1 shows
an example of such a potential.

It is not surprising that under these circumstances
we find
 bound states of negative energy, however, under appropriate
conditions, this
 potential can also display a localized state of zero energy.
Thus, it is again
 the physics of the upside--down pendulum that allows us to
understand this
 phenomenon.

We first concentrate on the one--dimensional case.
Guided by the analogy between the Newton and the Schr\"{o}dinger equation we
expect the corresponding ground state energy wave function to exhibit a slowly
varying envelope that is modulated with the period of the periodic potential;
but where does the binding come from? The origin of this bound state lies in
the modulation of the envelope---without the modulation there is no binding. 
The modulation can be easily understood: the particle is more likely to be in
a stable minimum of the potential rather than in an unstable maximum. 

The curvature of the wave function at a given position, expressed by its
second derivative, is determined,
not only by the potential, but also by the wave function at that point.
Indeed, according to the Schr\"{o}dinger equation (\ref{Schrodinger}),
it is the product of the
potential and the wave function that governs the curvature. This quantity
contains, apart from other terms, the product of the oscillatory potential and
the modulations of the wave function. The simplest case of a cosine
modulation gives rise to the square of the cosine, thus creating a constant
potential and a part that varies in space with twice the modulation frequency.

We focus on the constant part and note that this constant is always
negative. In order to explain this feature we recall that the modulation of
the envelope of the wave function is such that at the minima of the potential
we have local maxima of the wave function and vice versa. Hence, the
modulation and the potential defer in their sign and their product creating
the potential well is negative. The fact that the potential only wiggles over
a finite domain makes this constant negative potential into an attractive
potential well whose width is determined by the range of the oscillations. 
The depth of the well involves, apart from other parameters, the square of the
height of the periodic potential.
For more mathematical treatment and
explicit expressions for the well, we refer to the Appendix.

There is a close analogy between the ground state in the periodic potential
and the problem of the upside--down pendulum. In both cases the binding effect
is a consequence of the average of the product of two oscillatory functions:
for the pendulum it is the time dependent modulation of the foundation and the
time dependent angle of oscillation, which creates a harmonic oscillator of
positive steepness. When considering the particle in the periodic potential,
it is the product of the position dependent potential and the wave function
modulated in space that gives rise to a well. However, there is a dramatic
difference: whereas the modulation of the pendulum is on for all times, the
spatial modulation is confined to a certain domain of space. It is this
confinement in space, together with the creation of an effective potential,
which defines the bound state.

\section*{5. Zero-energy ground state}

The spatial domain, over which the wiggles are present, and the depth of the
potential well, determine the ground state energy.
Is it possible to choose these parameters so as to achieve
a ground state of zero energy?  The answer is: it depends! In one dimension
the answer is a flat no, but in three dimensions it is possible, as we shall
now show.

We start by considering the situation in one dimension.
As we decrease the depth of the well and the energy of the ground state
approaches zero, the tails of the wave function reach more and more into the
forbidden region outside of the well. This feature is a consequence of the
fact that the decay is governed by the square root of the absolute value of
the energy. In the limit of zero energy the wave function is no longer
localized; no mechanism prevents the wave from flooding all space.
Consequently, there is no ground state of zero energy in one dimension.  

Is it possible to construct such a zero-energy bound state in higher dimensions?
In order to answer this question, we first concentrate on the
three-dimensional case and consider a potential $V=V(r)$
that wiggles along the radial
direction. Beyond a final radius $r_{0}$, the potential vanishes. The
time independent Schr\"{o}dinger equation 

\begin{equation}
\frac{d^2 u_{0}(r)}{dr^2} + \frac{2M}{\hbar^2} \ [E- V(r)] \ u_{0}(r)=0
\label{radial}
\end{equation}

\noindent for the radial energy wave function $u_{0}=u_{0}(r)$
corresponding to vanishing angular momentum is then identical to the
one-dimensional case.  Due to the product of the oscillatory potential
 and the
wave function, and the finite domain of oscillations, we find an attractive
potential well of radius $r_{0}$. There is however, a subtle difference to the
one-dimensional case: it is the origin where the radial wave function has to
vanish \cite{rau} corresponding to an infinitely high potential wall at the
origin.  This feature is a consequence of the spherical symmetry. Therefore,
the infinite wall at the origin provides one side of the effective well
creating even a node of the wave function. The other wall of finite height is
due to the effective potential together with the finite range of the
oscillatory domain. 

For appropriate parameters such as the final radius and the depth of the
modulation of the potential we can fit a ground state of negative energy into
this well, as shown in Fig.2.
The corresponding wave function displays a node at the
origin and
 an exponential decay in the classically forbidden domain beyond the
critical
 radius, with a single maximum lying in between. Can we now take the
limit of
 zero energy and thus realize a zero-energy ground state?   

The answer is yes! Indeed, we can arrange the depth of the well such that
exactly one quarter of a period of a sine oscillation fits into the potential
well. Why a quarter of a period? Since the decay of the wave function outside
of the potential is governed by the square root of the energy we find that for
zero energy the wave function is constant. Due to the condition of continuity
on the wave function it must take on the value at the wall. The radial wave
function displays a node at the origin and increases like a sine function to a
constant value. However, the sine function must merge smoothly
into the constant, which is only possible at a point where the sine
function
 has an extremum. Since we want to have the ground state without a
node, except
 the one at the origin, the only possibility for such a merger
appears at one
 quarter of a period \cite{bri} of the sine function. 

Obviously, this zero--energy ground state radial wave function does not display
any localization. The localization becomes apparent when we recall that we
have to divide the radial wave function by the radial variable in order to
find the total wave function, that is

\begin{equation}
\psi_{0}(r) =\frac{u_{0}(r)}{r}.
\end{equation}
Indeed, now the total wave function enjoys a
maximum at the origin and decays with the inverse power of the radius. We
emphasize however, that the total wave function is not square integrable since
the volume element $4 \pi r^2 dr$ brings in the square of the radius. This contribution
cancels exactly the factor creating the localization in the probability
density, that is, in the absolute value squared of the wave function. For this
reason we refer to this state as localized state rather than bound state.  

We conclude this discussion by briefly addressing the case of positive energies.
Here, the exponential decay into the classically forbidden region of a wave
function corresponding to negative energy turns into right and left running
plane waves of positive energy. Obviously, plane waves are not localized wave
functions, nevertheless, the total wave function is concentrated at the origin
due to the inverse power of the radius; in this sense we have localized wave
functions in the continuum.

\section*{6. Quantum anti-centrifugal force}

The physics of a bound state in two dimensions is very different from the
one in other dimensions; and this for many reasons.
In \cite{cir} we have studied this case in more detail. In the
present paper we only highlight and motivate the results of \cite{cir}. 
For detailed derivations we refer to that article. To bring out the peculiarities
of the two--dimensional case we first consider the case of no external
potential. 
 
In two dimensions a vanishing angular momentum does not imply \cite{flu} a
vanishing centrifugal potential as in three dimensions.
Indeed here the radial Schr\"odinger equation reads

\begin{equation}
\frac{d^2 u_{0}(r)}{dr^2} + \frac{2M}{\hbar^2} \ [E-V_{Q}(r)] \ u_{0}(r)=0.
\label{polar}
\end{equation}

\noindent We can trace the
origin of the new potential 

\begin{equation}
 V_{Q}(r) \equiv  - \frac{\hbar^2}{2M} \ \frac{1}{4 r^2} 
\end{equation}

\noindent back to the non-vanishing commutation relation
between the operators of momentum and the radial unit vector. This potential
is, therefore, a true quantum potential. Since, in addition, it is attractive
rather than repulsive we have named it the quantum anti-centrifugal
potential \cite{cir}. Moreover, the potential is attractive only in a
two--dimensional world. For one or three dimensions it vanishes, but for
higher
 than three dimensions it becomes repulsive. 

The quantum anti-centrifugal potential has brought us into a rather unusual
situation. Usually, we perform wave mechanics in a classical external
potential. In the present context however, we have a quantum potential. This
feature stems from the reduction of space from three to two dimensions,
reminiscent of the Born-Oppenheimer approximation in molecular physics. A
molecule is a quantum system consisting of several degrees of freedom. In the
framework of the Born--Oppenheimer approximation we solve the Schr\"{o}dinger
equation for the electrons for a fixed position of the nuclei and then use the
resulting potential curves, that is, the electronic energies as a function of
the separation of the nuclei to determine their relative motion. In this sense
we indeed perform wave mechanics on potentials that are not due to external
forces but due to quantum mechanics. In the example of the nuclear motion in a
molecule, the appearance of wave mechanics in a quantum potential is a
consequence of the reduction of the degrees of freedom \cite{bri2}.

The attractive quantum anti-centrifugal potential manifests itself even in
the energy wave function of a free particle, that is in the absence of any
other external potential. The ordinary Bessel functions $J_{0}$ and
$N_{0}$ are two independent solutions of the time independent Schr\"{o}dinger
equation for positive energy.
Both show a bunching of nodes towards the
origin, as suggested by an attractive potential leading to an acceleration of
the particle towards the origin and demonstrated for the example
of $J_{0}$ in Fig.3.

We have to compare
and contrast this
 situation with the case of a particle with one unit of
angular momentum, where
 the centrifugal potential is indeed repulsive. It is
worth noting that the
 quantum anti-centrifugal potential reduces the strength
of repulsion of the
 centrifugal potential and the effective potential is not
as repulsive as in
 the classical case. In the case of one unit of angular
momentum, the
 solutions of the time independent Schr\"{o}dinger equation are 
the ordinary Bessel
functions $J_{1}$ and $N_{1}$. As expected by the repulsive potential
enforcing a deceleration of the particle as it approaches the origin, the
nodes of the Bessel functions show anti-bunching towards the origin. 

One might wonder if the quantum anti-centrifugal potential is strong enough
to support a bound state with negative energy. In the transition from
positive to negative energies real wave numbers transform into purely
imaginary ones as a consequence of the quadratic dispersion relation between
energy and wave number.
Therefore, the ordinary Bessel functions $J_{0}$ and $N_{0}$
turn into modified Bessel functions $I_{0}$ and $K_{0}$.
Since $I_{0}$ explodes for large arguments \cite{abr}, that is,
large radial distances, this solution does
not satisfy the requirement of exponential decay enforced by the attractive
quantum anti-centrifugal potential. Only the function $K_{0}$ achieves this
goal, but it has another disease: it explodes at the origin like a logarithm.  

This singularity at the origin indicates that the solution

\begin{equation}
u_{0}(r)=\sqrt{\frac{k}{\pi}}\sqrt{kr}K_{0}(kr)
\end{equation}
of the Schr\"odinger equation, shown in Fig.4,
does not describe a truly free particle but
that there is a delta function potential present. The strength of this
external potential
 determines \cite{wod} the energy eigenvalue of this bound
state. Moreover, due
 to the quantum anti-centrifugal potential the
probability of finding the
 particle within a given area in space is
concentrated along a band around the
 location of the delta function; a
remarkable feature unique to two dimensions.
 In a similar arrangement in one
or three dimensions the maximum of the
 probability is at the location of the
potential minimum.  This property might
 have interesting applications for
guiding atoms along wires \cite{den} or
 electro-magnetic waves in wave-guides
\cite{col}.
 
 We now return briefly to the case of an energy eigenstate in a
two--dimensional
 oscillatory potential of finite range. As emphasized in the
previous
 paragraph, the case of vanishing angular momentum contains the most
quantum
 effects. Here, the ground state wave function of negative energy
feels the
 combination of the quantum anti-centrifugal potential and the
potential well
 created by the product of the wave function and the potential.

\section*{7. Light induced potentials, Trojan wave packets and the lone
electron}

The examples discussed in the preceding sections represent but only a small
subsection of the class of unusual bound states. They have been selected
because they follow or are motivated by the physics of the upside-down
pendulum. However, there are many more intriguing bound states in atomic and
molecular physics \cite{gw}.  We conclude our paper by briefly summarizing
three
 examples starting with one that is still closely related to our main
theme of
 the upside-down pendulum and gradually moving away towards new
frontiers.     
 
Consider a diatomic molecule with various electronic states. Within the
Born--Oppenheimer approximation the electronic states provide the potentials
for the relative motion of the two nuclei. A strong external time dependent
laser field interacts with the electrons and couples the individual electronic
potential surfaces.  The wave functions for the vibratory motion of the nuclei
are therefore coupled through a time dependent periodic interaction
Hamiltonian. The product of the interaction Hamiltonian and the nuclear wave
functions govern the time evolution of the wave functions; a feature that is
again reminiscent of the upside-down pendulum. In the latter the product
consists of the time dependent steepness of the oscillator and the phase
angle.  The only difference lies in the fact that in the classical case the
Newton equation is of second order, whereas the Schr\"{o}dinger equation is a
first order.  However, this distinction is superficial since we are dealing
with the nuclear wave functions corresponding to different electronic states
and hence, with a system of differential equations of first order. 

Due to this product of two periodic functions with the same period (interaction
Hamiltonian and wave function) we find again a constant which lifts up the
electronic potential by integer multiples of the energy of the absorbed
photon. In mathematical terms this energy elevation is a consequence of
Floquet states. However, it can be viewed as just one more illustration of
the upside-down pendulum with counter-intuitive consequences. Indeed, when we
consider a binding and a repulsive potential this constant shift in energy can
result in a crossing of the potential curves. Since in a diatomic molecule
there exists a non-crossing rule we obtain an avoided crossing forming new
potentials; one is binding and one is repulsive. However, these potentials
are rather peculiar because each of them consists of parts of the two original
potentials. The formation of new bound states \cite{bes} in this light induced
molecule has been observed experimentally \cite{san}.

Trojan asteroids are celestial bodies that are held in their positions by the
gravitational forces of the sun and Jupiter. A similar phenomenon can occur
\cite{bia} in an atom where the electron plays the role of a Trojan asteroid,
and the nucleus substitutes for one of the two planets. A circularly polarized
electromagnetic field simulates the effect of the second planet. Indeed, the
electromagnetic interaction of the electron with the field and the Coulomb
attraction between the electron and the nucleus replace the gravitational
attraction of the celestial bodies. In a frame rotating with the
electromagnetic field, the motion of the electron is in a plane and governed
by three potentials which depend on the two-dimensional radial direction: the
Coulomb attraction of the nucleus, the centrifugal potential of the
circular motion and the linear potential arising from the interaction of the
electron with the electromagnetic field. For short distances the Coulomb
potential dominates, whereas for large distances the repulsive
potential of the circular motion prevails. Consequently, for intermediate
distances, an unstable potential maximum occurs. This feature is in
complete analogy with the upside--down pendulum. But where does the stabilizing
drive come from? The answer to this question lies in the fact that
the radial motion
is coupled to the angular motion. The latter is locked to the rotating
electromagnetic field and the electron
performs angular vibrations around this stable point
of equilibrium. These vibrations couple into the radial motion creating a
dynamically binding potential. This classical picture also holds true
for the quantum case as verified by extensive studies of wave packets
propagating in this minimum.  

In our last example of unusual bound states it is the balance of
static forces that is responsible for the formation of a bound state of a
lone electron in a Rydberg atom in crossed electric and magnetic fields
\cite{fau}. The crossed-field situation is of particular appeal in the field
of quantum chaology since in this system energy is the only conserved
quantity. The classical system displays chaos and is therefore of interest in
the search for fingerprints of chaos in the corresponding quantum system. 
Apart from these questions of quantum chaos there is another quite interesting
aspect of this system. The electron in the atom experiences, not only the
attractive Coulomb potential of the nucleus, but also a linear potential due
to a constant electric field and a binding harmonic oscillator potential due
to the magnetic field.  The superposition of all three potentials creates a
local potential minimum far away from the nucleus.  An electron bound in this
minimum displays a large dipole moment that can be observed by sending the
atom through an inhomogeneous electric field. Indeed, the experiments
\cite{den} have confirmed the existence of this far outside lying minimum.

We have come a long
way on our journey into unusual bound states. Starting from the classical
physics of the upside-down pendulum we have been led to the phenomenon of a
zero-energy ground state in a periodic potential in three dimensions.  The
two-dimensional world has even more surprises in store: the attractive quantum
anti-centrifugal force that manifests itself in a free particle through the
bunching of the nodes or even in a bound state. Here the probability of
finding the particle is concentrated in the domain where no force is acting. 
This feature resulting from the action of the quantum anti-centrifugal
potential could be useful in guiding atoms along wires or electromagnetic
waves along fibers. Experiments on trapping cold atoms by single photons
\cite{hoo} or along a whispering gallery mode of a glass sphere resonator
\cite{ver} are yet more realizations of the physics of the upside-down pendulum
and give us confidence that our predictions of unusual bound and localized
states can be verified experimentally in the near future.    

\section*{Acknowledgments}

This work started when I (WPS) had the great privilege
to be a postdoc with
John A. Wheeler more than a decade ago.
Since that time, John and I have
frequently returned to the topics addressed in the present paper.
These discussions took place in trains, planes and
automobiles and at various locations such as
Hightstown, High Island, and Ulm. I thank John
for this wonderful time, his outstanding hospitality
at Hightstown and High Island and especially for
the unique experience
to work with him. The present paper is
partially based on notes and sketches of figures prepared
jointly.
I am grateful for many fruitful discussions and John's deep
insights and, in particular, for allowing us to use the material
that had originally
been obtained in close collaboration with him.
The proceedings of the Lake Garda Conference are a most
welcome opportunity to finally summarize this project started
many years ago. Moreover, we are grateful to I. Bia\l ynicki-Birula
for a critical reading of the manuscript.
Two of us (MAC and WPS) thank R. Bonifacio for his hospitality and for
organizing
a most interesting conference in the splendid surroundings of Lake
Garda. The work of WPS is partially supported by the Deutsche
Forschungsgemeinschaft,
moreover, he gratefully acknowledges a travel grant
from the Universit\"{a}t
Ulm which made part of this research possible.

\section*{Appendix}

In contrast to the main body of the paper, we pursue in the present
Appendix a more mathematical approach.
In particular, we emphasize the similarities between the classical equation
of motion for the upside--down pendulum and the Schr\"{o}dinger equation
for a nonrelativistic particle in a periodic potential. For this purpose,
we first cast the corresponding equations into a dimensionless form.
This approach allows us to simultaneously derive equations for the
dynamics contained in the
macro-- and micro--motion of the pendulum or the envelope function and
the modulation of the ground state wave function.

We start by summarizing the classical equation of motion for the
upside--down pendulum. For the sake of simplicity we
consider the limit of small angles $\varphi$ measured relative
to the vertical position.

A vertical acceleration $\ddot{f}$ of the foundation translates into a vertical
acceleration of the mass $M$ of the pendulum and adds to the gravitational
acceleration $g$ pointing downwards. Here, the sign of $\ddot{f}$ is
crucial: when $\ddot{f}$ is positive, that is, when the pendulum is
accelerated upwards, the gravitational acceleration pointing
downwards is reduced. Likewise, when $\ddot{f}$ is negative, that is, when
the pendulum is accelerated downwards, the gravitational
acceleration is increased. Consequently, the total acceleration is
$g-\ddot{f}$. Hence for small angles $\varphi$ the force
tangentially to the rod of the pendulum of length $L$ reads

\begin{equation}
M L \frac{d^2\varphi}{dt^2} = M \left( g-\ddot{f}\: \right) \varphi,
\end{equation}
giving rise to

\begin{equation}
\frac{d^2\varphi(t)}{dt^2}+\left[-\Omega^2+\frac{1}{L}\ddot{f}
\right] \varphi(t)=0, \end{equation}
where $\Omega^2\equiv g/L$.

It is reasonable to assume that the modulation
$f$ is a periodic function of period $T\equiv 2\pi /
\nu$. This time scale allows us to
introduce a dimensionless variable $\theta \equiv
\nu t$ and the differential equation of the upside--down
pendulum takes the form

\begin{equation}
\frac{d^2\varphi}{d\theta^2} + \left[ \kappa
+a(\theta)\right]\varphi=0.
\label{eight}
\end{equation}
Here we have introduced the abbreviations
$\kappa\equiv -(\Omega/\nu)^2$ and
$a(\theta)\equiv f''(\theta)/L$ for the steepness
of the inverted harmonic oscillator and the acceleration due to the
modulation, respectively.
Moreover, prime denotes differentiation with
respect to $\theta$.

We now make the connection to the time independent
Schr\"{o}dinger equation

\begin{equation}
\frac{d^2 u(x)}{dx^2}+\frac{2M}{\hbar^2}\left[E-V(x)\right] u(x)=0
\label{sch1}
\end{equation}
for a particle of mass $M$ in the potential $V(x)$,
which is periodic in space with period $\lambda\equiv 2\pi / k$.

When we introduce the recoil energy $E_{r}\equiv (\hbar k)^2/(2M)$ and
the dimensionless coordinate $\theta\equiv kx$, the Schr\"{o}dinger
equation takes the form

\begin{equation}
\frac{d^2 u}{d\theta^2}+\left[ \eta-v(\theta) \right]
u=0
\label{ten}
\end{equation}
where the dimensionless energy eigenvalue $\eta\equiv E/E_{r}$ and the
potential
$v(\theta)\equiv V(\theta/k)/E_{r}$ are scaled in units of the recoil energy.

The two dimensionless equations of the driven pendulum
(\ref{eight}) and the quantum particle in the periodic potential (\ref{ten})
have identical structure. In Secs.2 and 4
we have presented qualitative arguments
to explain the motion of the pendulum as a
superposition of a slow motion (macromotion) and a rapid motion (micromotion)
or, in the language of wave functions, to decompose
the energy wave function into an envelope and a modulation.
We now take advantage of the analogy between the two systems
to support these qualitative arguments
by rigorous mathematics. For this purpose we start from the equation

\begin{equation}
u''(\theta)+\left[\eta-v(\theta)\right] u(\theta)=0.
\label{schacc}
\end{equation}

Again, we emphasize that $u$ can either be the phase angle
of the pendulum or the wave function. Likewise, $\theta$ represents
time or position and the drive $v(\theta)$ is either the acceleration
of the foundation or the potential. 

In order to find the solution of (\ref{schacc}), we make the {\em
Ansatz}

\begin{equation}
u(\theta)={\cal A}(\theta)\left[ 1-\epsilon v(\theta) \right],
\label{ansatz}
\end{equation}
consisting of
the product of the slowly varying envelope $\cal{A}(\theta)$ and the
modulation caused by the drive $v(\theta)$.

We substitute the {\em Ansatz} (\ref{ansatz}) into
(\ref{schacc}), which yields

\begin{eqnarray}
& & {\cal A}''+\left[ \eta+\epsilon v^2\right]{\cal A}
\nonumber \\
& & -\left[ {\cal A}''+\left( \eta +\frac{1}{\epsilon}+
\frac{v''}{v}\right) {\cal A}\right] \epsilon v \nonumber \\
& & -2\epsilon {\cal A}' v'=0.
\label{star}
\end{eqnarray}

So far, our analysis is exact.
We now solve the equation (\ref{star}) in an approximate way.
For this purpose, we first neglect the last contribution in (\ref{star})
involving the product of the first derivatives of the envelope function
and the potential. This approximation is
justified since $v'$ is out of phase with $v$ and cannot therefore
lead to a significant contribution. Moreover, we choose $\epsilon$
such that 

\begin{equation}
\frac{1}{\epsilon}+\frac{v''}{v}=0.
\end{equation}
We emphasize that this is not possible in a strict sense, since $v''/v$
is not necessarily a constant. However, in the oscillatory domain of the
potential the most elementary model
$v(\theta)\propto \cos(\theta)$ suggests the estimate $v''/v\simeq
-1$ and hence we find

\begin{equation}
\epsilon=+1.
\end{equation}

Provided the envelope ${\cal A}$ satisfies the equation

\begin{equation}
{\cal A}''+\left[ \eta+\epsilon v^2\right]{\cal A}=0
\end{equation}
also the second contribution in Eq.(\ref{star}) vanishes to this order
in $\epsilon$. This is due to the fact that the second contribution is already
multiplied by $\epsilon$.

For the case of the
upside--down pendulum, the parameter $\eta$ is negative. In the
absence of the drive, the negative value of $\eta$
leads to an exponential growth of ${\cal A}$. However,
the square of the drive is always positive and can create
an overall positive coefficient in front of ${\cal A}$,
providing an oscillation rather that an exponential explosion.

In case of the quantum particle, it is advantageous to
write the equation for ${\cal A}$ in the form of a
Schr\"{o}dinger equation

\begin{equation}
{\cal A}''(\theta)+\left[ \eta-v_{eff}(\theta)\right]
{\cal A}(\theta)=0.
\end{equation}
Here we have averaged the square of the potential over
a period of the oscillations giving rise to a smooth
effective potential

\begin{equation}
v_{eff}(\theta)\equiv -\frac{1}{2\pi}\int_{\theta}^{\theta+2\pi}
d\theta' v^2(\theta').
\end{equation}

We note that this potential is always negative.
When the modulation of the potential is
confined in space, for example by a window function $\mu(\theta)$,
and the potential $v(\theta)$ is of the form

\begin{equation}
v(\theta)\approx \mu(\theta)C(\theta)
\end{equation}
where $C$ is a periodic function, we can approximate
the effective potential by a potential well

\begin{equation}
v_{eff}(\theta)=-\frac{1}{2} \mu^2(\theta).
\label{well}
\end{equation}
In this case, the envelope ${\cal A}$ is an energy
wave function of this potential well and the parameter
$\eta$ is the dimensionless energy eigenvalue.
Hence, $\eta$ is not determined from the
outside, but rather by the well (\ref{well}) itself. If the well
is deep enough, it can support, apart from the ground state, 
also excited states with negative energy.

\begin{figure}
\caption{Formation
of a bound state in the accordion potential.  This one-dimensional potential
oscillates in a confined region of space and decays rapidly to a constant
outside, therefore being reminiscent of an accordion. The modulations in the
wave function and in the potential (top) give rise to an effective potential
well (bottom) determining the envelope of the bound state.}
\end{figure}

\begin{figure}
\caption{Formation of a bound or localized state
in a three-dimensional accordion potential. We consider the cases of the
eigenenergy being smaller (left column), equal (middle column) or larger (right
column) than the free space potential energy. The modulation of the potential
enforces a modulation of the radial wave function (top row). In complete
analogy to the one-dimensional case, the envelope of the wave function follows
from the effective potential well (bottom row) created by the product of the
modulations in the wave function and the potential. For negative energies the
radial wave function decays exponentially in the classically  forbidden regime
whereas for vanishing energy it is constant. Positive energies lead to
oscillatory wave functions. In the last two cases the radial wave function is
not localized. However, the probability density $|\psi_{0}(r)|^2=
|u_{0}(r)/r|^2$ shown in the middle row is localized due to the additional
factor $1/r^2$.}
\end{figure}

\begin{figure}
\caption{Node bunching and anti-bunching of energy eigenfunctions
of a free particle in a two-dimensional space.
The centrifugal potential corresponding
to a non-vanishing angular momentum is repulsive (bottom inset). This feature
gives rise to an anti-bunching of the nodes of the energy eigenfunction
$u_{1}(\rho) \equiv \sqrt{\rho} J_{1}(\rho)$, that is an anti-bunching of the
zeros $j_{1,n}$ of the Bessel function $J_{1}$. In contrast, the centrifugal
potential corresponding to a vanishing angular momentum is attractive (top
inset). This feature leads to a bunching of the nodes of the wave function
$u_{0}(\rho) \equiv \sqrt{\rho} J_{0}(\rho) $, that is a bunching of the zeros
$j_{0,n}$ of the Bessel function $J_{0}$. As a measure $g_{m}(n) \equiv \pi /
\Delta_{m}(n)$ of bunching or anti-bunching we use the inverse of the
separation $ \Delta_{m}(n) \equiv j_{m,n+1} - j_{m,n} $ of neighbouring zeros
$j_{m,n}$ of the $m$-th Bessel function $J_{m}$ in units of the free space
separation $\pi$.  Squares correspond to $g_{1}(n)$ of a weakly repulsive
centrifugal potential. Diamonds represent $g_{0}(n)$ with an attractive
centrifugal potential--the   quantum anti-centrifugal potential. The physics
of the non-relativistic free particle does not contain an
intrinsic unit of length. When
we define a dimensionless  length $\rho \equiv k r$ where $k$ is the wave
number, the dimensionless energy eigenvalue is unity.}  
\end{figure}

\begin{figure}
\caption{Bound state of a `` free '' particle. The attractive
quantum anti-centrifugal potential $V_{Q}$ supports a single bound state of
negative dimensionless energy. The corresponding energy wave function
$u_{0}(\rho) \equiv \sqrt{\rho / \pi} K_{0}(\rho)$ is normalizable, since it
decays exponentially into the classically forbidden regime provided by the
quantum anti-centrifugal potential. Here $K_{0}$ is the Neumann function of
purely imaginary argument. At the origin the wave function has infinite
steepness indicating the presence of an additional potential located solely at
the origin, that is a delta function potential.}
\end{figure}


\begin{references}
\bibitem{bet2} J. A. Wheeler in many discussions with one of us
(WPS) repeatedly credited H. Bethe with the discovery of the problems
leading to the principle of strong focusing.
Unfortunately, we have not been able
to locate the appropriate reference. We have talked to Prof. Bethe,
but also he was not able to point out the article.
\bibitem{bri}
A similar reasoning appears \cite{whe} in general relativity, that is in
geometrodynamics, when we determine the metric coefficients of a
system. They follow from Brill's equation \cite{whe}, which is similar to
the time independent Schr\"{o}dinger equation for zero energy. In contrast to
quantum mechanics where the Schr\"{o}dinger equation determines the energy
eigenvalues we now have to solve the equation under the constraints that the
 wave is not allowed to have nodes and corresponds to zero energy.
\end{references}
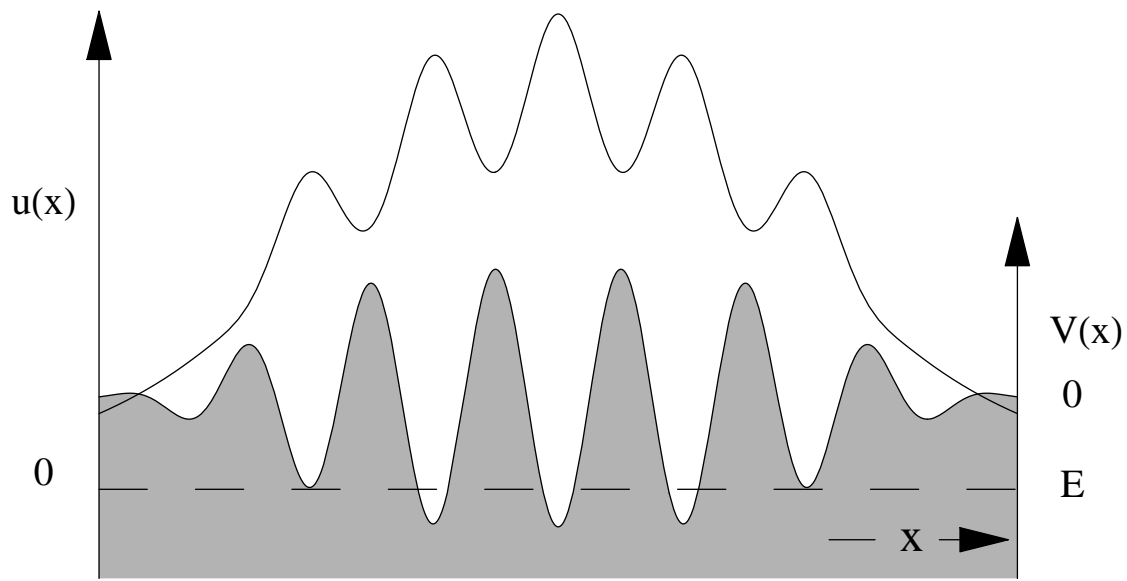
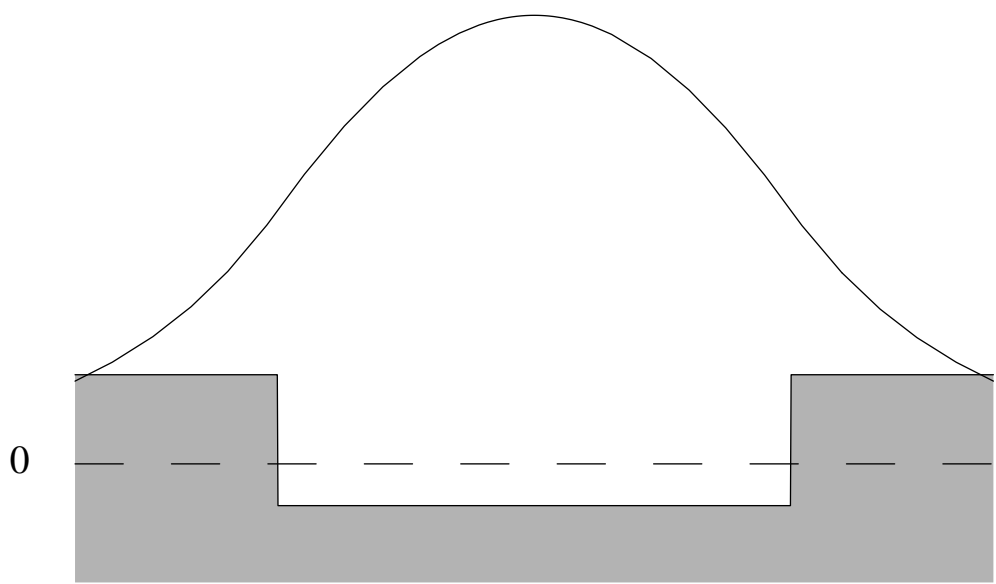
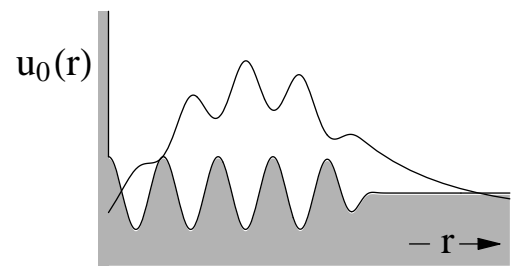
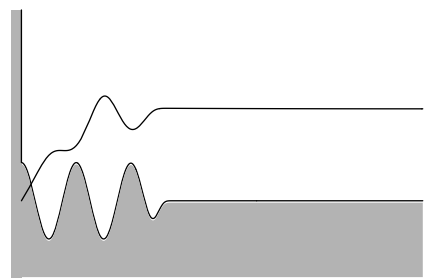
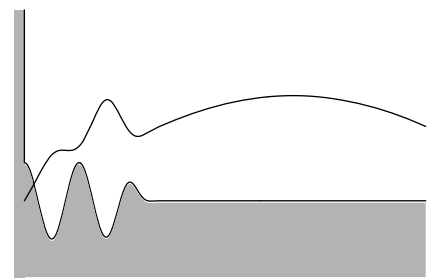
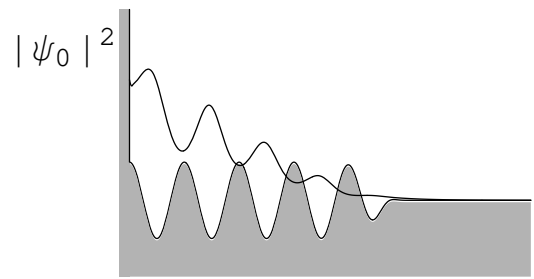
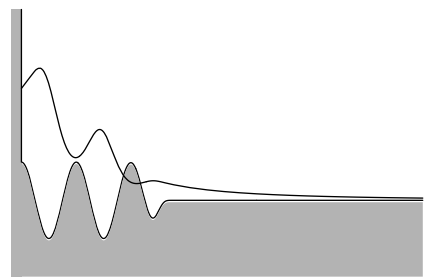
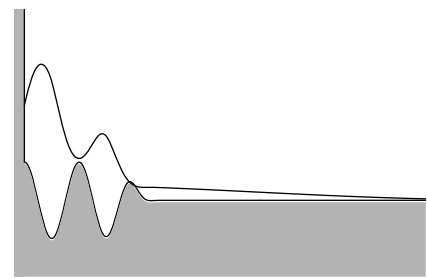
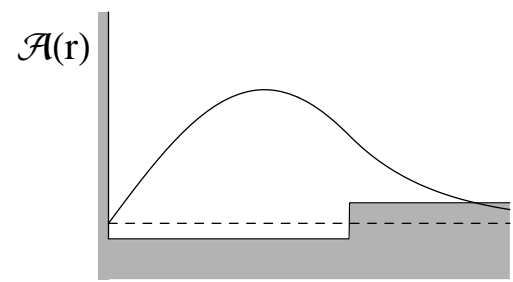
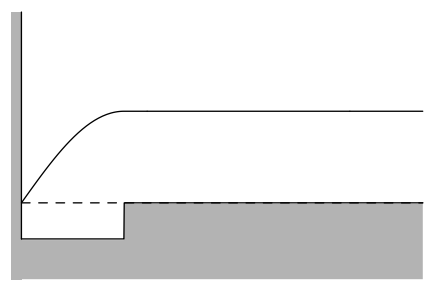
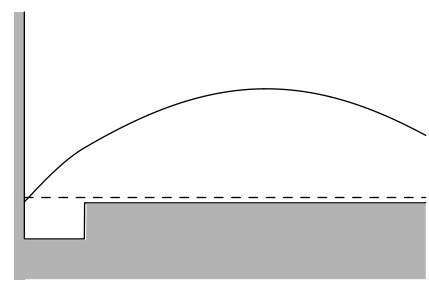
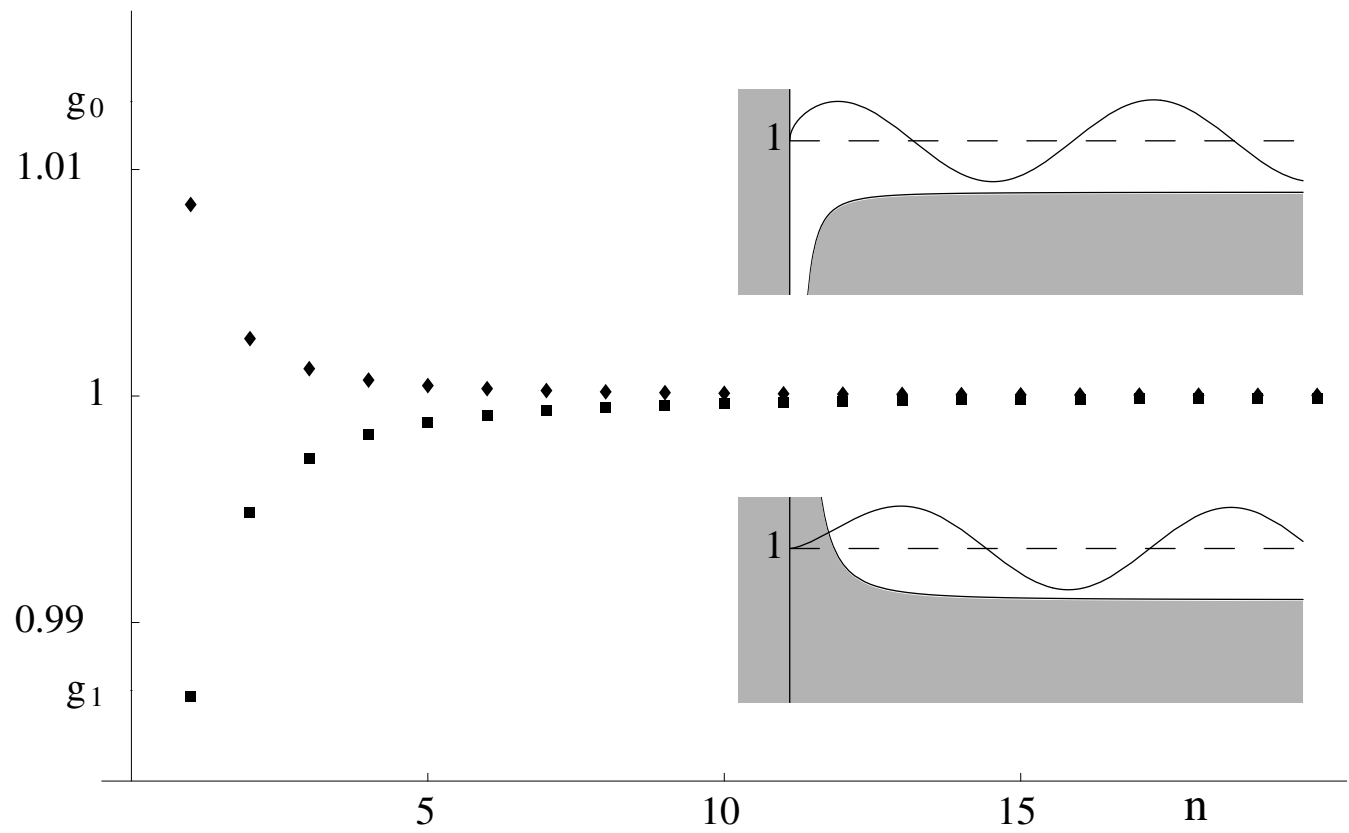
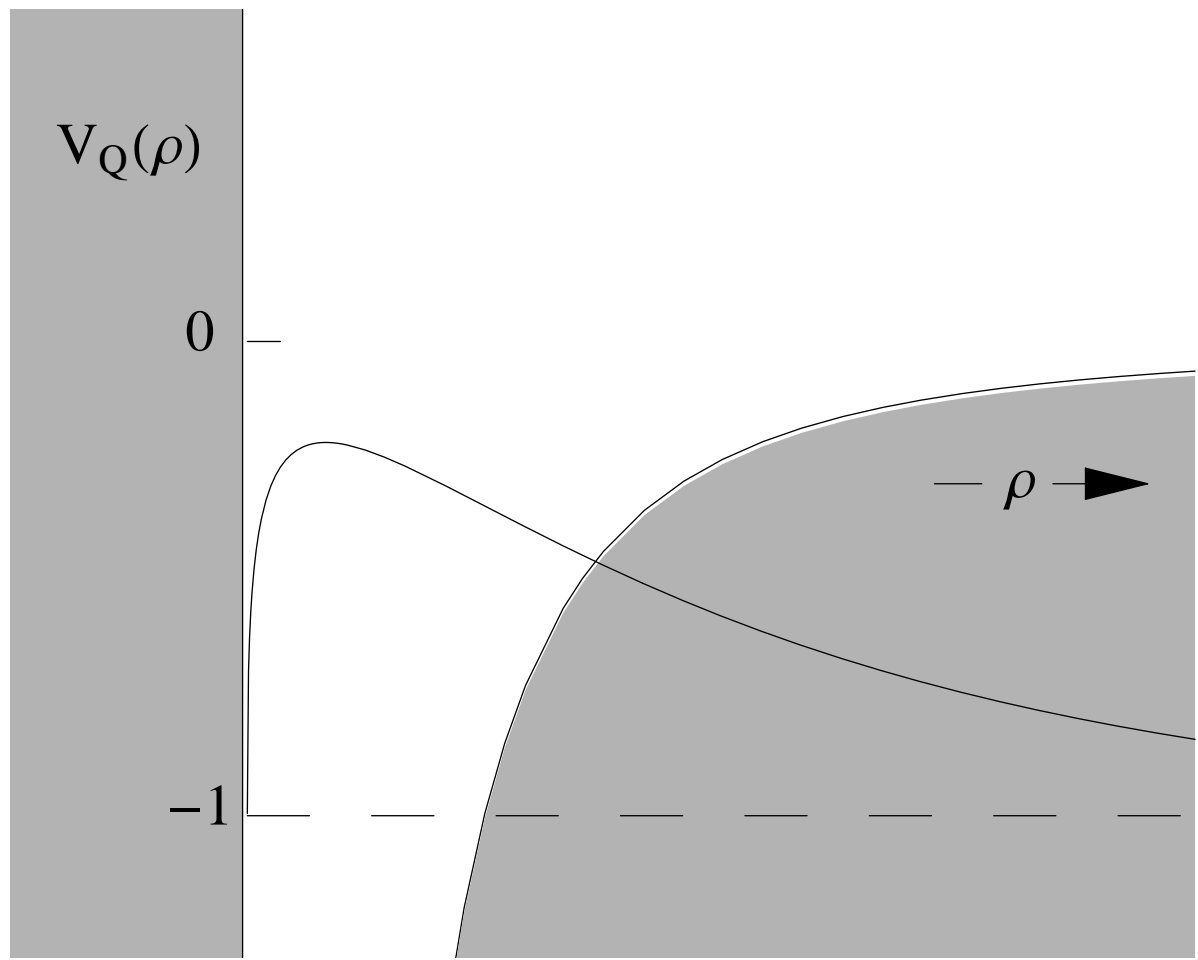
\end{document}